\documentclass{svproc}
\usepackage{url}

\usepackage{amsmath,graphicx}
\begin{document}
\mainmatter              
\title{Lattice QCD Study of Doubly Heavy Bottom Tetraquarks}
\titlerunning{Doubly Heavy Tetraquarks}  
%
\author{Bhabani Sankar Tripathy\inst{1,2} \and Nilmani Mathur\inst{3} \and 
M. Padmanath\inst{1,2}}
\authorrunning{B. S. Tripathy et al.} 
%
\tocauthor{Bhabani Sankar Tripathy, Nilmani Mathur, and M. Padmanath}
\institute{
The Institute of Mathematical Sciences, CIT Campus, Chennai, 600113, India \\
\email{bhabanist@imsc.res.in} \and 
Homi Bhabha National Institute, Training School Complex, Anushaktinagar, Mumbai, 400094, India \and 
Department of Theoretical Physics, Tata Institute of Fundamental Research, Mumbai, 400005, India
}
\maketitle

\begin{abstract}
Hadrons, composed of quarks and gluons bound by Quantum Chromodynamics (QCD), traditionally classified as baryons (three quarks) and mesons (quark-antiquark pairs). Nothing in the theory of QCD stands against the existence of exotic hadrons with more complex quark contents. Recent discoveries by LHCb and Belle, such as X, Y, Z states and $T_{cc}(3875)$, have renewed interest in these states. Understanding the binding mechanism within of these exotic states provides insights into QCD's non-perturbative dynamics. This work presents lattice QCD studies of two-meson interactions, involving bottom quarks, on MILC ensembles, exploring heavy tetraquark channels.

\end{abstract}
\section{Introduction}
Over the last two decades, there has been significant progress in exotic hadron spectroscopy, both theoretically and experimentally Ref.\cite{Lebed:2016hpi,Guo:2017jvc,Bicudo:2022cqi}. To understand QCD in the low-energy regime, lattice QCD provides a first-principles approach. In this work, we focus on tetraquarks containing bottom quarks, specifically those with the valence structure $bb\bar{u}\bar{d}$($T_{bb}$) in the $I(J^P) = 0(1^+)$ quantum channel. In the heavy quark limit, this system is expected to form a bound state.
Numerous past studies, including phenomenological approaches, diquark models, and QCD sum rules, suggest the existence of such a bound state \cite{Lebed:2016hpi,Guo:2017jvc,Bicudo:2022cqi}. More recently, lattice QCD studies, including our earlier work in Ref.~\cite{Junnarkar:2018twb}, indicate the possibility of a bound state \cite{Francis:2024fwf}. This study improves upon our previous work \cite{Junnarkar:2018twb} by incorporating multiple volumes and a box-sink smearing setup to better control systematic uncertainties. Additionally, we perform a scattering amplitude analysis, enabling a more rigorous determination of the binding energy of $T_{bb}$. \vspace{-0.2cm}

 \section{Lattice setup}
We utilize a set of four lattice QCD ensembles with $N_f=2+1+1$ dynamical flavors generated by the MILC Collaboration~\cite{MILC:2012znn}, employing the Highly Improved Staggered Quark (HISQ) action. These ensembles feature two different lattice volumes and three lattice spacing ($a$), with the finest lattice spacing being $a \sim 0.0582$ fm. The strange and charm quark masses in the sea are tuned to their respective physical values, while the dynamical light quark masses are isosymmetric ($m_u = m_d$) and heavier than their physical values. A summary of our setup is presented in Table 1 of Ref.~\cite{Tripathy:2025vao}. 

For valence quarks up to the strange quark mass, we employ the overlap fermion action, while for the bottom quark, a non-relativistic QCD action is used. The tuning of the bottom-quark mass on each ensemble follows Fermilab's prescription, based on the kinetic mass of spin-averaged 1S bottomonia. This is expressed as $a\bar{M}^{\bar{b}b}_{\text{kin}} = \frac{3}{4}aM_{\Upsilon} + \frac{1}{4}aM^{\bar{b}b}_{\eta_b}$, which is determined on each ensemble. The strange quark mass is tuned by equating it to the mass of a hypothetical pseudoscalar $\bar{s}s$ meson of 688.5 MeV. 

We consider four light pseudoscalar meson masses: 0.5, 0.6, 0.7, and 3.0 GeV. Among these, the quark propagators corresponding to $M_{ps} \sim 0.7$ GeV and 3.0 GeV are at the physical strange and charm quarks, respectively. Finite-volume spectra are systematically evaluated for all ensembles and pseudoscalar meson masses. This is followed by an amplitude analysis of the scattering of mesons near the nearest decay threshold, such as $B$ and $B^{*}$ in the case of $bb\bar{u}\bar{d}$. To suppress higher momentum modes, wall-source smearing is employed, and box-sink correlators are utilized to accurately identify the large time saturation of ground state energy in the correlation functions. 
\vspace{-0.2cm}

\section{Finite Volume and Amplitude Analysis}
We determined the finite-volume spectrum by examining the Euclidean two-point correlation function, which is expressed as:
\begin{equation}
    \mathcal{C}_{ij}(t) = \sum_{\mathbf{x}} \left<\Phi_i(\mathbf{x},t)\tilde{\Phi}_j^{\dagger}(0)\right> = \sum_n \frac{\mathcal{Z}_i^n \tilde{\mathcal{Z}}_j^{n\dagger}}{2E^n} e^{-E^n t}.
\end{equation}
Here, $\Phi_i(\mathbf{x},t)$'s are interpolating operators belonging to the $T_{1g}$ finite-volume irrep, which are the finite-volume counterparts of $J^P = 1^+$ in infinite volume. For $bb\bar{u}\bar{d}$, we utilized both diquark-antidiquark ($\Phi_\mathcal{D}$) and meson-meson ($\Phi_{\mathcal{M}_{BB^*}}$) type operators, given as:
\begin{align}
\Phi_{\mathcal{M}_{BB^*}}(x) &= \left[\bar{u}(x)\gamma_i b(x)\right] \left[\bar{d}(x)\gamma_5 b(x)\right]
            - \left[\bar{u}(x)\gamma_5 b(x)\right] \left[\bar{d}(x)\gamma_i b(x)\right], \nonumber \\
\Phi_\mathcal{D}(x) &= \big[\left(\bar{u}(x)^T C\gamma_5 \bar{d}(x) - \bar{d}(x)^T C\gamma_5 \bar{u}(x)\right) \times \left(b^T(x) C\gamma_i b(x)\right)\big].
\label{eq:bb1}
\end{align}
Here, $C = i\gamma_y\gamma_t$ represents the charge conjugation matrix. The objects inside the square brackets are color neutral. The lowest relevant two-body scattering channel in this case corresponds to $BB^*$, while the lowest inelastic scattering channel is $B^*B^*$. The lowest three-body scattering channel, $BB\pi$, is sufficiently high, and any associated left-hand cuts \cite{Du:2023hlu} are beyond the scope of this work.

To extract the energy levels of the low-lying spectrum, we analyze the correlation matrices variationally by solving a generalized eigenvalue problem (GEVP): 
\begin{equation}
\mathcal{C}(t)v^{(n)}(t) = \lambda^{(n)}(t) \mathcal{C}(t_0)v^{(n)}(t).
\label{gevp}
\end{equation}
The extraction of the energy spectrum involves fitting the eigenvalue $\lambda^{(0)}(t)$ or the ratios $R^0(t) = \lambda^{(0)}(t)/{\mathcal{C}_{m_1}(t) \mathcal{C}_{m_2}(t)}$ to their large-time exponential behavior. Here, $\mathcal{C}_{m_i}(t)$ represents the two-point correlation function for the meson $m_i$. Our final results are based on fitting the ratio correlators, which effectively suppresses systematic uncertainties.
\begin{figure}[htbp]
    \centering
    \begin{minipage}[t]{0.48\linewidth}
        \centering
        \includegraphics[width=1.0\linewidth, height=0.5\linewidth]{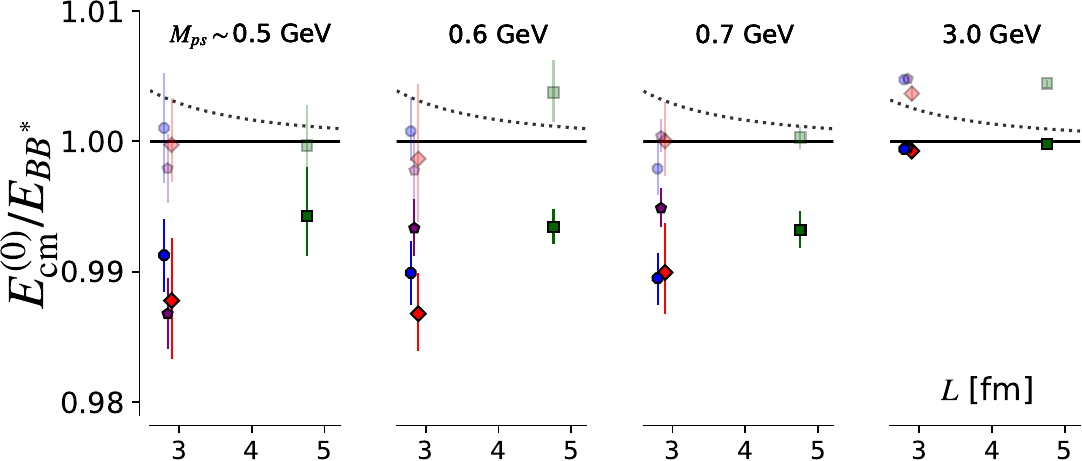}
        \label{nrqcd_bb}
    \end{minipage}%
    \hfill
    \begin{minipage}[t]{0.48\linewidth}
        \centering
        \includegraphics[width=1.0\linewidth, height=0.5\linewidth]{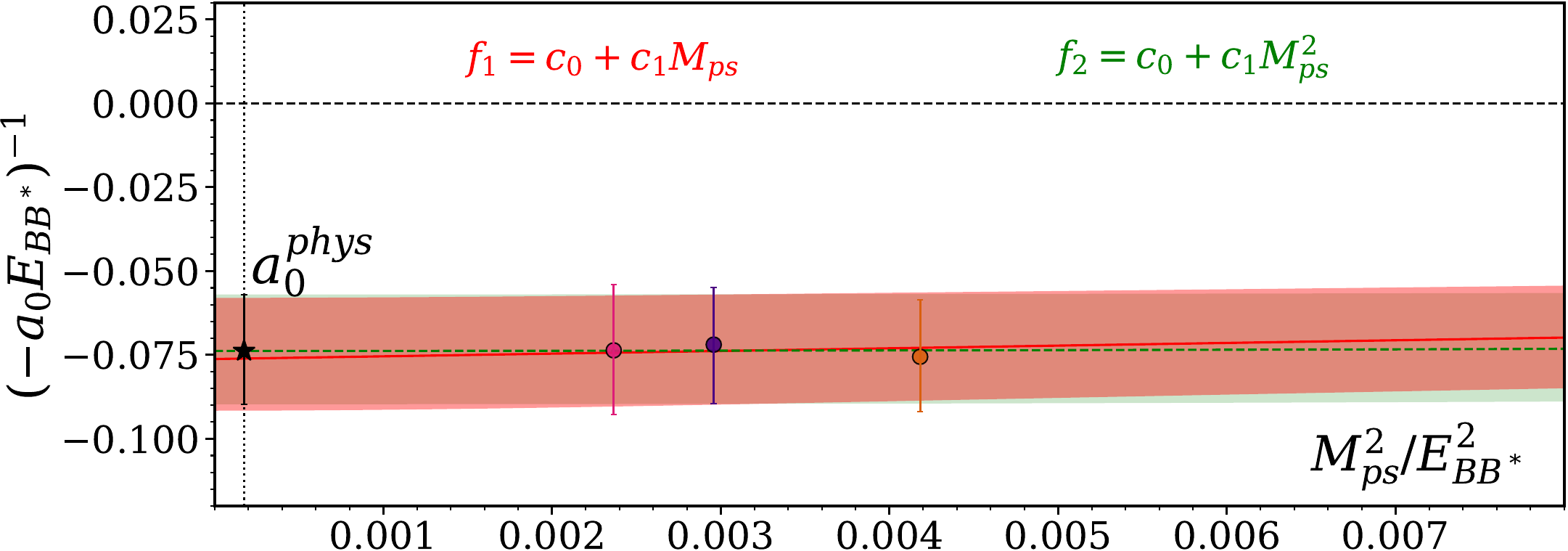}
        \label{other_plot}
    \end{minipage}
    \caption{(Left) Ground state energies in units of the $BB^{*}$ threshold energy on all ensembles and for all $M_{ps}$. (Right) Continuum extrapolated $p \cot \delta_0$ estimates of the $BB^{*}$ system as a function of $M^2_{ps}$ in units of $E_{BB^{*}}$.}
    \label{fig:combined_plots}
\end{figure}
In the left panel of Figure~\ref{fig:combined_plots}, we present the ground state energy estimates relative to the nearest two-body decay threshold $BB^{*}$ across various $M_{ps}$ and ensembles, following the marker and color conventions of Ref. \cite{Tripathy:2025vao}. Faded markers represent first excited state spectrum and lowest omitted noninteracting level with one momenta(in lattice unit) is shown as dotted curve. The energy estimates from the eigenvalue correlators include additive offsets from the dynamics of the NRQCD, which are corrected in the final results based on eigenenergy splittings $\Delta E$ from ratio correlators $R^0(t)$. The corrected meson masses are determined as $\tilde{M}_{B^{(*)}} = M_{B^{(*)}}$ $- 0.5\overline M^{\bar bb}_{lat} + 0.5 \overline M^{\bar bb}_{phys}$, where $\overline M^{\bar bb}_{lat}(\overline M^{\bar bb}_{phys})$ are the spin-averaged bottomonium masses $1S$ of the lattice (experiment). Negative energy shifts in the ground states indicate attractive \( B \)-\( B^* \) interactions, whereas the faded markers for the first excited states align with the elastic threshold.

\underline{Amplitude analysis:} We extract the \( S \)-wave scattering amplitudes under elastic approximations using the finite-volume spectrum quantization formalism by Lüscher and its extensions~\cite{Luscher:1990ux,Briceno:2014oea}. The \( S \)-wave phase shift, \(\delta_0(p)\), is obtained from finite-volume energy levels using $pcot[\delta_0(p)] = 2Z_{00}\left[1;\left(\frac{pL}{2\pi}\right)^2\right]/(L\sqrt{\pi})$
where $p$ is the center-of-momentum frame scattering momentum, related to the total energy $E_{cm}$ through $4sp^2 = (s-(M_{1}+M_{2})^2)(s-(M_{1}-M_{2})^2)$
with  $s = E_{\text{cm}}^2$. 
The energy dependence of the relevant scattering amplitude is determined by enforcing the quantization condition, following Appendix B of Ref.~\cite{Padmanath:2022cvl}. We then analyze the amplitude for near-threshold poles indicating bound states or resonances, where a bound state is signaled by  $p\cot\delta_0 = +\sqrt{-p^2}$. We parametrize the elastic two body threshold scattering amplitude with a term related to scattering length and lattice spacing dependence($f=A^{[0]}+ A^{[1]} \cdot a$). By fitting across all $M_{ps}$, followed by chiral extrapolation, we find the scattering length at $M_{ps} =M^{phy}_{ps}$ as $a^{phy}_0=0.25(^{4}_{3}) ~fm$ which corresponds to binding energy $-116(^{+30}_{-36})$ MeV shown in the right panel of Figure~\ref{fig:combined_plots}.
\vspace{-0.2cm}

 \section{Conclusion}
 We conducted a lattice QCD study that focused on the $BB^*$ scattering with $I(J^P)=0(1^{+})$ quantum numbers. We observe a negative shift in the ground state energy  with respect to the $BB^*$ threshold. The lattice-extracted amplitude suggests the exitence of a bound state with binding  energy $-116(^{+30}_{-36})$ MeV. \\ 
 
\noindent\underline{Acknowledgments}: This work is supported by the Department of Atomic Energy, Government of India, under Project Identification Number RTI 4002. Computations were carried out on the Cray-XC30 of ILGTI, TIFR (which has recently been closed), and the computing clusters at the Department of Theoretical Physics, TIFR, Mumbai, and IMSc Chennai. MP gratefully acknowledges support from the Department of Science and Technology, India, SERB Start-up Research Grant No.SRG/2023/001235.

%
%


\begin{thebibliography}{6}

\bibitem{Lebed:2016hpi}
R.~F.~Lebed, R.~E.~Mitchell and E.~S.~Swanson,
Prog. Part. Nucl. Phys. \textbf{93}, 143-194 (2017)
doi:10.1016/j.ppnp.2016.11.003
[arXiv:1610.04528 [hep-ph]].

\bibitem{Guo:2017jvc}
F.~K.~Guo, C.~Hanhart, U.~G.~Mei\ss{}ner, Q.~Wang, Q.~Zhao and B.~S.~Zou,
Rev. Mod. Phys. \textbf{90}, no.1, 015004 (2018)
[erratum: Rev. Mod. Phys. \textbf{94}, no.2, 029901 (2022)]
doi:10.1103/RevModPhys.90.015004
[arXiv:1705.00141 [hep-ph]].

\bibitem{Bicudo:2022cqi}
P.~Bicudo,
Phys. Rept. \textbf{1039}, 1-49 (2023)
doi:10.1016/j.physrep.2023.10.001
[arXiv:2212.07793 [hep-lat]].

%
\bibitem{Francis:2024fwf}
A.~Francis,
Prog. Part. Nucl. Phys. \textbf{140}, 104143 (2025)
doi:10.1016/j.ppnp.2024.104143

\bibitem{Junnarkar:2018twb}
P.~Junnarkar, N.~Mathur and M.~Padmanath,
Phys. Rev. D \textbf{99}, no.3, 034507 (2019)
doi:10.1103/PhysRevD.99.034507
[arXiv:1810.12285 [hep-lat]].

\bibitem{MILC:2012znn}
A.~Bazavov \textit{et al.} [MILC],
Phys. Rev. D \textbf{87}, no.5, 054505 (2013)
doi:10.1103/PhysRevD.87.054505
[arXiv:1212.4768 [hep-lat]].

\bibitem{Tripathy:2025vao}
B.~S.~Tripathy, N.~Mathur and M.~Padmanath,
[arXiv:2503.09760 [hep-lat]].

\bibitem{Du:2023hlu}
M.~L.~Du, A.~Filin, V.~Baru, X.~K.~Dong, E.~Epelbaum, F.~K.~Guo, C.~Hanhart, A.~Nefediev, J.~Nieves and Q.~Wang,
Phys. Rev. Lett. \textbf{131}, no.13, 131903 (2023)
doi:10.1103/PhysRevLett.131.131903
[arXiv:2303.09441 [hep-ph]].

\bibitem{Meinel:2022lzo}
S.~Meinel, M.~Pflaumer and M.~Wagner,
Phys. Rev. D \textbf{106}, no.3, 034507 (2022)
doi:10.1103/PhysRevD.106.034507
[arXiv:2205.13982 [hep-lat]].


\bibitem{Michael:1985ne}
C.~Michael,
Nucl. Phys. B \textbf{259}, 58-76 (1985)
doi:10.1016/0550-3213(85)90297-4
\bibitem{Luscher:1990ux}
M.~Luscher,
Nucl. Phys. B \textbf{354}, 531-578 (1991)
doi:10.1016/0550-3213(91)90366-6

\bibitem{Briceno:2014oea}
R.~A.~Briceno,
Phys. Rev. D \textbf{89}, no.7, 074507 (2014)
doi:10.1103/PhysRevD.89.074507
[arXiv:1401.3312 [hep-lat]].

\bibitem{Padmanath:2022cvl}
M.~Padmanath and S.~Prelovsek,
Phys. Rev. Lett. \textbf{129}, no.3, 032002 (2022)
doi:10.1103/PhysRevLett.129.032002
[arXiv:2202.10110 [hep-lat]].
\end{thebibliography}
\end{document}